 \newcommand{\CIV}{\ion{C}{iv}}
 
 \newcommand{\Hb}{\mbox{H$\beta$}}
 \newcommand{\Ha}{\mbox{H$\alpha$}}
 
 \newcommand{\Lya}{\mbox{Ly$\alpha$}}
 \newcommand{\kms}{\mbox{$\mathrm{km\,s}^{-1}$}}

\documentclass[letter]{aa}  

\usepackage{graphicx}
\usepackage{txfonts}
\begin{document} 

   \title{The Close AGN Reference Survey (CARS)}

   \subtitle{What is causing Mrk~1018's return to the shadows after 30 years?\thanks{Based on Cycle 17 DDT program (ID: 18789, PI: G. Tremblay) approved by the 
{\it Chandra} Director, Dr.~Belinda Wilkes. Based on Cycle 23 DDT project with the NASA/ESA Hubble Space 
Telescope (ID: 14486, PI: B. Husemann) approved by {\it HST} Director Dr. Kenneth Sembach.}}
   \titlerunning{What is causing Mrk~1018's return to the shadows after 30 years?}
   \authorrunning{B. Husemann}
   \author{B. Husemann$^{1,\star}$,
	  T. Urrutia \inst{2}, 
	  G. R. Tremblay$^{3,\dagger}$,
	  M. Krumpe \inst{2},
	  J. Dexter \inst{4},
	  G. Busch \inst{5}, 
	  F. Combes \inst{6}, 
	  S. M. Croom \inst{7,8}, 
	 \\ T. A. Davis \inst{9}, 
	  A. Eckart\inst{5,10},
          R. E. McElroy \inst{7,8},  
          M. Perez-Torres \inst{11,12},
          M. Powell \inst{3}, 
          J. Scharw\"achter \inst{13}
          }

   \institute{European Southern Observatory, Karl-Schwarzschild-Str. 2, D-85748 Garching b. M\"unchen, Germany, $\star$ ESO Fellow\\
              \email{bhuseman@eso.org}
              \and Leibniz-Institut f\"ur Astrophysik Potsdam (AIP), An der Sternwarte 16, D-14482 Potsdam, Germany 
              \and Yale Center for Astronomy and Astrophysics, Yale University, 52 Hillhouse Ave., New Haven, CT 06511, USA, $\dagger$ Einstein Fellow
              \and Max-Planck-Institut f\"ur extraterrestische Physik, Giessenbachstr. 1, D-85748 Garching, Germany
              \and I. Physikalisches Institut, Universit\"at zu K\"oln, Z\"ulpicher Stra{\ss}e 77, D-50937 K\"oln, Germany 
              \and LERMA, Observatoire de Paris, College de France, PSL, CNRS, Sorbonne Univ., UPMC, F-75014 Paris, France
              \and Sydney Institute for Astronomy, School of Physics, University of Sydney, NSW 2006, Australia
	      \and ARC Centre of Excellence for All-sky Astrophysics (CAASTRO) 
              \and School of Physics \&\ Astronomy, Cardiff University, Queens Buildings, The Parade, Cardiff, CF24 3AA, UK
              \and Max-Planck-Institut f\"ur Radioastronomie, Auf dem H\"ugel 69, D-53121 Bonn, Germany
	      \and Instituto de Astrof\'isica de Andaluc\'ia, Glorieta de las Astronom\'ia, s/n, E-18008 Granada, Spain
	      \and Visiting Scientist: Departamento de F\'isica Teorica, Facultad de Ciencias, Universidad de Zaragoza, Spain
	      \and Gemini Observatory, Northern Operations Center, 670 N. A'ohoku Pl., Hilo, Hawaii, 96720, USA
             }

   \date{}


 \abstract{We recently discovered that the active galactic nucleus (AGN) of Mrk~1018 has changed optical type again after 30 years as a type 1 AGN. Here we combine  
\textit{Chandra}, \textit{NuStar}, \textit{Swift}, \textit{Hubble Space Telescope} and ground-based observations to explore the cause of this change. The 2--10\,keV flux declines by a factor 
of $\sim$8 between 2010 and 2016. We show with our X-ray observation that this is not caused by varying neutral hydrogen absorption along the line-of-sight up to the 
Compton-thick level. The optical-UV spectral energy distributions are well fit with a standard geometrically thin optically thick accretion disc model that seems to obey the expected $L\sim T^4$ 
relation. It confirms that a decline in accretion disc luminosity is the primary origin for the type change. We detect a new narrow-line absorber in \Lya\ blue-shifted by $\sim$700\,\kms\ with respect 
to the systemic velocity of the galaxy. This new \Lya\ absorber could be evidence for the onset of an outflow or a companion black hole with associated gas that could be related to the 
accretion rate change.  However, the low column density of the absorber means that it is not the direct cause for Mrk 1018's changing-look nature.}
  

   \maketitle
%

\section{Introduction}
Active galactic nuclei (AGN) and some X-ray binaries and  are thought to be powered by accretion of material onto a black hole (BH). They commonly show significant variability at optical-to-X-ray 
wavelengths on short timescales; this can be well described by noise processes \citep[e.g.][]{Nandra:1997,McHardy:2004,Mushotzky:2011}. The variability timescale is expected to scale with the mass 
of the BH and is therefore longest for super-massive BHs (SMBH) and can reach up to several hundred years \citep[e.g.][]{McHardy:2006}. It is therefore difficult or even impossible to directly 
measure the long-term high-amplitude fluctuations of AGN over the required timescales.

Dramatic changes in the soft X-ray brightness or the strength of broad Balmer lines emitted in the broad line region (BLR) have been reported in some AGN. Examples of such 
a ''changing-look'' AGN caused by absorbing clouds passing in front of the nucleus and/or variable reflection components are NGC~4151 \citep[e.g.][and references therein]{Puccetti:2007},
NGC1365 \citep[e.g.][]{Risaliti:2009} and NGC~4051 \citep{Guainazzi:1998}, but these cloud events can be more common \citep{Markowitz:2014}. Prominent examples of AGN with 
appearing BLR are Mrk~1018 \citep{Cohen:1986}, NGC~1097 \citep{Storchi-Bergmann:1993}, and NGC~2617 \citep{Shappee:2014}, and examples with disappearing BLR are NGC 7603 \citep{Tohline:1976}, 
Mrk~590 \citep{Denney:2014}, SDSS~J0159+0033 \citep{LaMassa:2015}, and SDSS~J1011+5442 \citep{Runnoe:2016}. These events can either be explained by flares from tidal disruption events (TDEs) that 
are due to accretion of a star \citep[e.g.][]{Komossa:1999,Halpern:2004,Merloni:2015}, or intrinsic changes in the accretion disc flow depending on their light curves.

\citet{McElroy:2016} (hereafter Paper I) reported the surprising discovery that \object{Mrk 1018} ($z=0.035$), which turned from a type 1.9 to a bright type 1 AGN around 1984, has changed back 
to a type 1.9 nucleus after about 30 years. The optical continuum brightness dropped by an order of magnitude between 2010 and early 2016. While we discuss in Paper I that a TDE probably cannot 
explain the variability of Mrk~1018, several other options including a cloud event still appeared possible. In this Letter, we present follow-up Director's Discretionary Time (DDT) and archival X-ray 
and UV spectroscopic data that show that the changing classification is driven by accretion rate changes and not by an obscuration event.


\section{Observations and results}

\subsection{Chandra and NuStar X-ray spectroscopy}

\begin{figure}
   \centering
   \includegraphics[width=\hsize]{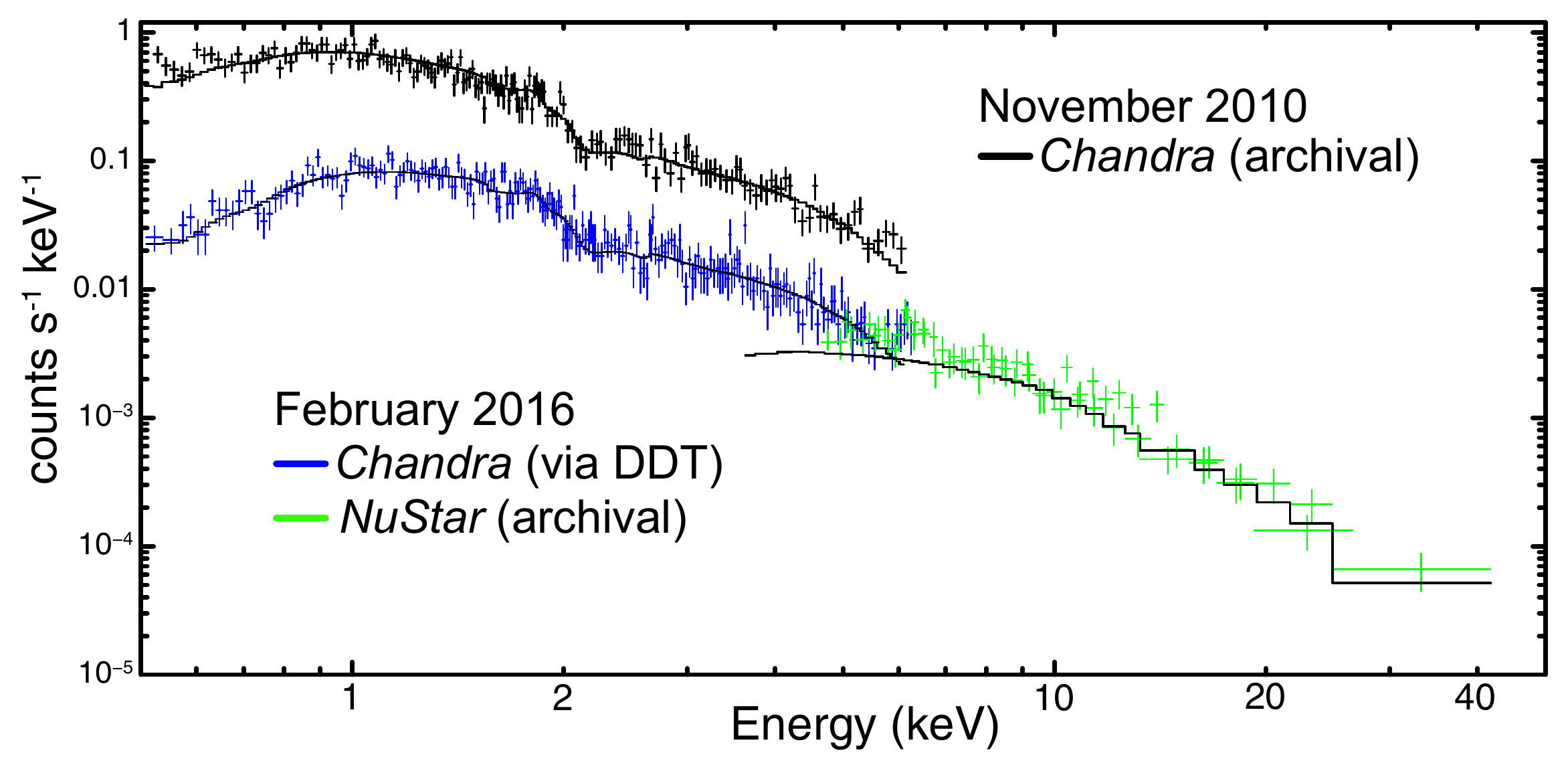}
      \caption{Comparison of the X-ray spectrum of Mrk~1018 in November 2010 and February 2016. For all data, we show the best-fit model of a simple 
power law plus Galactic absorption as a solid black line. The apparent mismatch near 5 keV between the 2016 {\it Chandra} and 2016 {\it NuStar} are instrumental effects included in the model. The 
X-ray flux dropped by a factor of 7.6 and both spectra are consistent with no $N_H$ absorption.}
   \label{fig:Chandra_spectra}
   \end{figure}

\textit{Chandra} observed Mrk~1018 on 2016-02-18 as part of a DDT request. It was observed on the nominal aimpoint of the back-illuminated ACIS-S3 chip for 27.2\,ksec. Mrk~1018 was also targeted on 
2016-02-10 with \textit{NuStar} \citep{Harrison:2013} for 21.6\,ksec as part of the public shallow Extragalactic Survey. The combined \textit{Chandra} and \textit{NuStar} X-ray spectrum covering 
0.5-50\,keV is shown in Fig.~\ref{fig:Chandra_spectra} together with an archival \textit{Chandra} observation taken on 2010-09-27  (ID: 12868, PI: Mushotzky).

Both \textit{Chandra} spectra were extracted with the \textit{CIAO} package (v4.5) and the latest \textit{CALDB} files (4.7.0) using standard settings for point sources. Similar settings for 
point sources were also employed for the \textit{NuStar} spectral extraction using the \textit{nupipeline}. Pile-up is severe only for the Chandra data from 2010 with a likelihood of $>$10\% 
in the brightest 9 pixels associated with the point spread function (PSF). To mitigate the effect of pile-up, we excluded these pixels when we extracted the spectrum and fitted a model. Afterwards 
we corrected for the loss of photons by fixing the model except for the normalization, which was then determined from the total spectrum in the 4--8\,keV range. All spectra were grouped with a 
minimum binning of 20 counts. We fitted an intrinsically absorbed power law together with absorption by the Galactic neutral hydrogen \citep[$N_{HI,Gal} = 2.43 \times 
10^{20}\,\mathrm{cm}^{-2}$,][]{Kalberla:2005} to the data. The \textit{Chandra} and \textit{NuStar} spectra from 2016 were fitted simultaneously. 

The 2010 and 2016 spectra and their fits are both consistent with no absorption beyond Galactic, in particular, even partial Compton-thick absorption is ruled out with {\it NuStar} in 2016. The 
best-fit model parameters and errors on the photon index $(\Gamma)$ and 2--10\,keV flux are listed in Table~\ref{table:XRT}. The relative normalization of the X-ray spectra indicates 
that the flux has dropped by a factor of 7.6 in 2016 compared to 2010. Furthermore, there appears to be a hint of a weak Fe K$\alpha$ line in the 2016 data.

\subsection{Swift X-ray monitoring}
While \textit{Chandra} obtained spectra with very high S/N of Mrk~1018, they only probe two epochs. The X-Ray Telescope (XRT) onboard the \textit{Swift} satellite has targeted Mrk~1018 
several times between 2005 and 2016 (see Table~\ref{table:XRT}). With these data we infer the evolution of X-ray brightness, $N_H$, and the $\Gamma$ based on interactive \textit{Swift} data 
processing and analysis pipeline (http://swift.asdc.asi.it).

Given the individual exposure times of up to a few ksec, only a simple spectral model consisting of a power-law plus Galactic absorption component is fitted. This simple model agrees well even with 
the higher S/N \textit{Chandra} observation. Pile-up does not significantly affect the \textit{Swift} data because of its wider PSF and can be ignored. We list the best-fit 
power-law value and the 2--10 keV flux in Table~\ref{table:XRT}. Considering the large uncertainties of the \textit{Swift}-based quantities from February 2016, the measurements broadly agree with our 
DDT \textit{Chandra} observation that has a much higher S/N around the same time.

\begin{table}
\caption{X-ray observations and analysis results}             
\label{table:XRT}      
\centering  
\begin{small}
\begin{tabular}{cccccc}        
\hline\hline                 
Date\tablefootmark{a} & $t_\mathrm{exp}$\tablefootmark{b} & $\theta_\mathrm{off}$\tablefootmark{c}  & $N_\mathrm{bin}$($\chi_\nu^2)$\tablefootmark{d} & $\Gamma$\tablefootmark{e} & 
$f_\mathrm{2-10\,keV}$\tablefootmark{f} \\ \hline  
 
53587(S) & 5.2 &   2.6 & 113(1.1) & $1.93\pm0.05$ & $1.11\pm0.08$\\
54271(S) & 3.3 &   4.4 & 58(1.1) & $1.91\pm0.08$ & $0.92\pm0.10$\\
54273(S) & 3.5 &   6.3 & 61(1.2) & $1.95\pm0.08$ & $0.78\pm0.07$\\
54275(S) & 4.1 &   5.9 & 73(1.0) & $1.95\pm0.07$ & $0.85\pm0.07$\\
54628(S) & 4.8 &   1.2 & 81(1.0) & $1.76\pm0.06$ & $0.97\pm0.08$\\
55527(C) & 22.7 & 0.0 & 169(1.2) & $1.68\pm0.04$ & $0.92\pm0.02$\\   
56450(S) & 1.3 &   2.0 & 14(1.1) & $1.42\pm0.18$ & $0.79\pm0.16$\\
56817(S) & 2.1 &   4.5 & 3(0.5) & $1.50\pm0.60$ & $0.16\pm0.09$\\
57429(S) & 3.7 &   3.7 & 9(1.3) & $1.75\pm0.27$ & $0.15\pm0.05$\\ 
57434(S) & 3.1 &   3.9 & 8(0.5) & $1.33\pm0.26$ & $0.25\pm0.08$\\    
57436(CN) & 27.2 & 0.0 & 511(1.1) & $1.62\pm0.03 $ &  $0.12\pm0.01$\\
\hline                                   
\end{tabular}
\end{small}
\tablefoot{
\tablefoottext{a}{Modified Julian date of observations with facility indicated in brackets C-\textit{Chandra}, S-\textit{Swift}, N-\textit{NuStar}}
\tablefoottext{b}{effective exposure time in ksec}
\tablefoottext{c}{off axis-angle in arcmin}
\tablefoottext{d}{Number of bins used for X-ray fitting and reduced $\chi^2$ of the best-fit model in brackets}
\tablefoottext{e}{Photon index with 90\% uncertainty range}
\tablefoottext{f}{Physical flux (Galactic-absorption corrected, on-axis corrected) between 2--10\,keV in units of $10^{-11}\,\mathrm{erg}\,\mathrm{cm}^{-2}\,\mathrm{s}^{-1}$}
}
\end{table}

\begin{figure}
   \centering
   \includegraphics[width=\hsize]{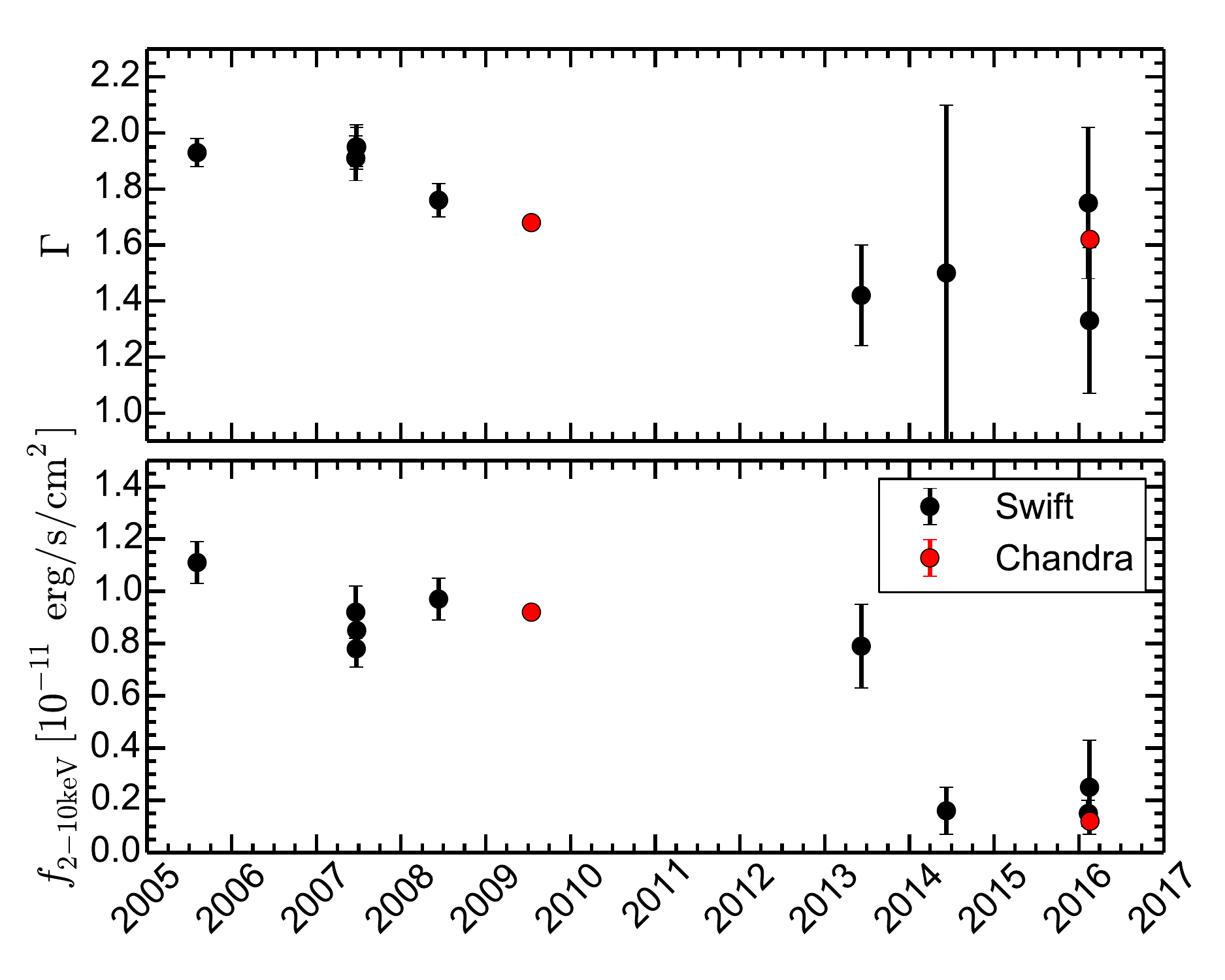}
      \caption{Time evolution of the X-ray photon index $\Gamma$ and the 2--10\,keV flux from 2005 until 2016 based on the \textit{Swift} and \textit{Chandra} data.}
         \label{fig:Xray_evolution}
   \end{figure}

In all \textit{Swift} observations with more than 2 ksec, a second fit with a free Galactic absorption parameter leads to lower values or within the 90\% uncertainty range of the 
Galactic-absorption value. Hence, during 2005 and 2016 the \textit{Swift} data show no signs of additional absorption along the line of sight, in agreement with the {\it Chandra} observations. 
We also find some variations in $\Gamma$ ranging between $\Gamma\sim1.9$ and $\Gamma\sim1.6$ during the period covered by \textit{Swift} observations (Fig.~\ref{fig:Xray_evolution}).

\subsection{HST FUV spectroscopy}
\begin{figure}
   \centering
   \includegraphics[width=\hsize]{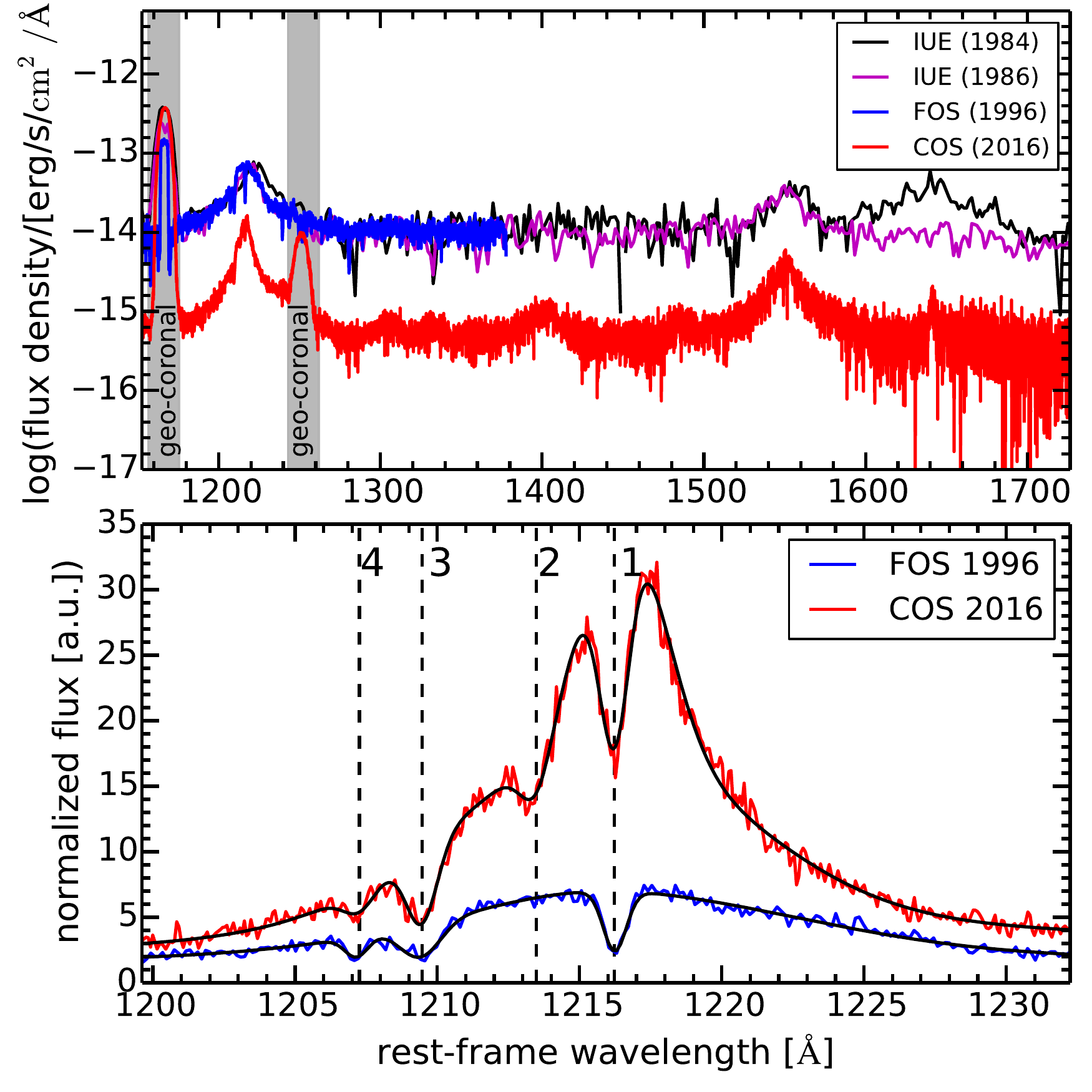}
      \caption{\textit{Upper panel:} FUV spectra of Mrk~1018 taken between 1984 and 2016 with IUE and the \textit{HST} FOS and COS. \textit{Lower panel:} Comparison of the Ly$\alpha$ 
emission-line shape in 1996 and 2016. The spectra are normalized so that the adjacent continuum level is one. The solid black lines are the best-fit model as described in the text.}
         \label{fig:FUV_spectra}
   \end{figure}

Hubble Space Telescope (\textit{HST}) FUV spectroscopy of Mrk~1018 was obtained with the Cosmic Origin Spectrograph \citep[COS,][]{Green:2012} on February 27 2016 for two orbits granted as 
DDT. The G140L grism provides a wavelength range covering \Lya\ and \CIV. A NUV acquisition image was also taken with the 
Primary Science Aperture and MIRRORB. The COS spectrum (Fig.~\ref{fig:FUV_spectra}) is compared to archival \textit{IUE} spectra from 1984/86 and \textit{HST} spectra 
taken with the Faint Object Spectrograph (FOS) in 1996. 

All spectra in the bright phase exhibit a FUV continuum flux density of ($1.3\pm0.3)\times10^{-14}\,\mathrm{erg}\,\mathrm{s}^{-1}\,\mathrm{cm}^{-2}\,\mathrm{\AA}^{-1}$ and  
$7.8\times10^{-16}\,\mathrm{erg}\,\mathrm{s}^{-1}\,\mathrm{cm}^{-2}\,\mathrm{\AA}^{-1}$ seen by COS, which is a factor of $\sim$17 fainter. The \Lya\ line is shown in the lower panel of  
Fig.~\ref{fig:FUV_spectra} after normalizing the adjacent continuum level to 1. The broad line is well modelled with three Gaussians plus single Gaussians for each 
absorption line. We measure a total Ly$\alpha$ flux of $16\times10^{-13}\,\mathrm{erg}\,\mathrm{s}^{-1}\,\mathrm{cm}^{-2}$ and 
$2.2\times10^{-13}\,\mathrm{erg}\,\mathrm{s}^{-1}\,\mathrm{cm}^{-2}$, respectively, which is a factor of $\sim$7 brighter than in 1996. The Ly$\alpha$ line width becomes narrower from 
$4170\pm62$\,\kms\ to $1330\pm122$\,\kms\ in FWHM. This is conceptually consistent with the resonant nature of the line. The \Lya\ photons produced in the past 30\,years are continuously absorbed and 
re-emitted and thereby able to scatter to larger distance with more quiescent kinematics. This may also explain why the broad \Lya\ line appears more symmetric than the asymmetric Balmer lines, 
as reported in Paper I.

Three \Lya\ narrow absorption lines (NALs) can be identified in the FOS spectrum taken in 1996. Surprisingly, an additional \Lya\ NAL appears 20 years later. We label the NALs from 1 to 
4 (Fig.~\ref{fig:FUV_spectra}).  NAL 1 has an equivalent width (EW) of 0.65\,\AA\ exactly at the systemic redshift of Mrk~1018. The new NAL 2 is blue-shifted by $\sim$700\,\kms\ with respect to the 
systemic redshift and has an EW of 0.32\AA. Absorbers 3+4 are blue-shifted by more than 1500\,\kms\ with EWs of 0.9\AA\ and 0.4\AA\ and less variable.  These EWs imply 
$N_H<10^{19}\,\mathrm{cm}^{-2}$ and do not lead to measurable X-ray absorption. Only absorber 1 is detected in \CIV, while absorber 2 remains undetected. The lower EW of absorber 2 combined with the 
lower S/N at \CIV\ during the fading phase may simply prevent a detection. Although \CIV\ is undetected for the new NAL, the short-time variability of the line at this strength can only be produced 
by neutral gas within the host galaxy. 

\section{Discussion}
\subsection{Inconsistency with a cloud event}
We have peculated in Paper I whether a dense cloud might be moving into our line of sight and causes the dimming of the nucleus. Such a scenario could explain the potential periodicity of such an 
event. For a BH mass of $M_\mathrm{BH}\sim8\times 10^7 M_{\sun}$ (Paper I) an orbital period of 30\,yr would correspond to a velocity of 3300\,\kms\ at a mean distance of 0.03\,pc. The combined 
high-quality \textit{Chandra} and \textit{NuStar} spectrum clearly shows that this scenario can be reliably ruled out because no significant $N_H$ can be detected. In particular, the 
high energies probed by \textit{NuStar} imply that Compton-thick obscuration is excluded. The apparent time evolution in $\Gamma$ and flux without any significant X-ray 
absorbing $N_H$ column density favours a scenario in which the physical state of 
the accretion disc underwent a significant change or reconfiguration. 

\subsection{Accretion disc changes probed by the spectral energy distribution}
\begin{figure}
   \centering
   \includegraphics[width=\hsize]{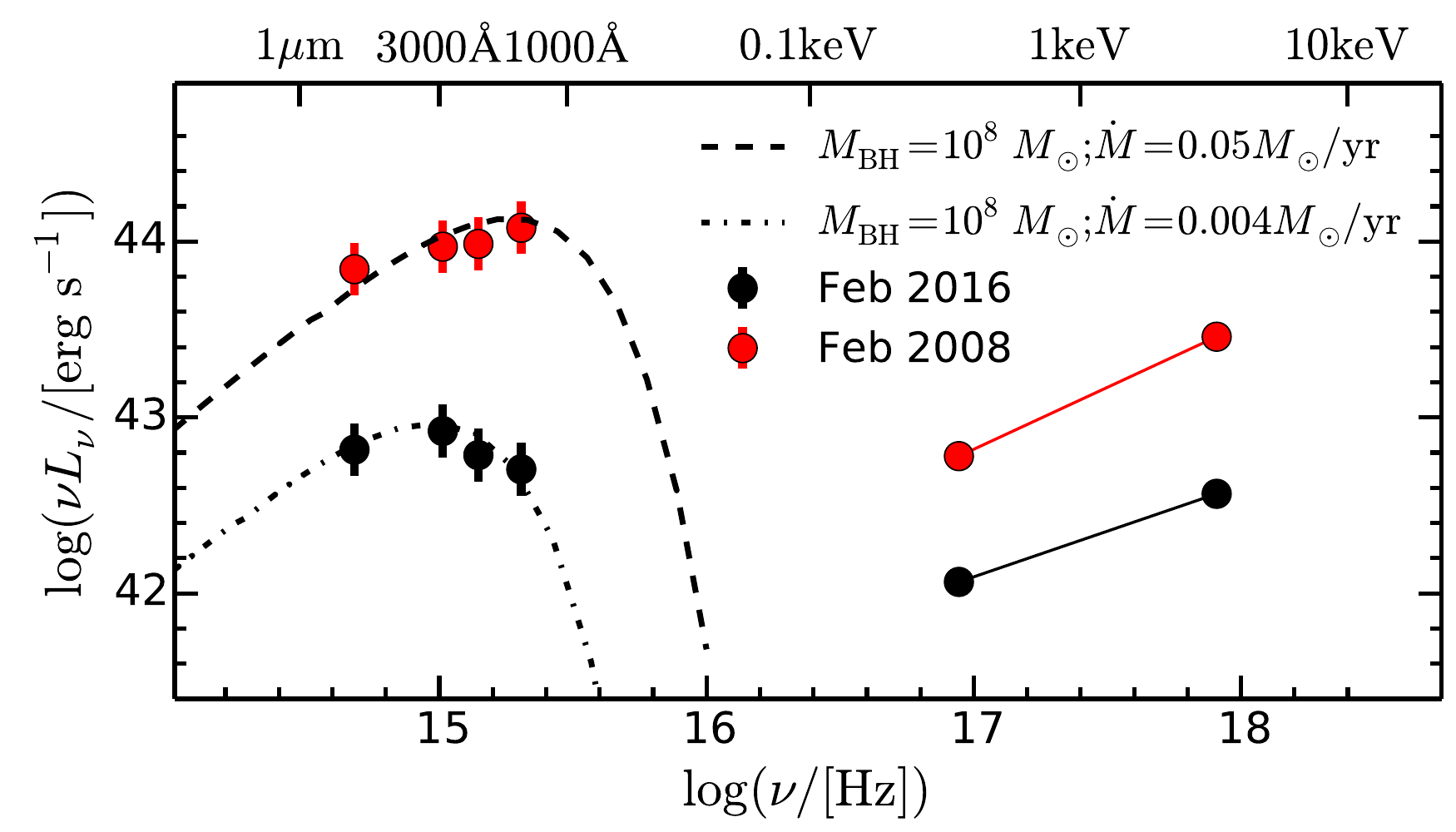}
      \caption{Optical to X-ray SED for Mrk~1018 for two epochs. Shown are the photometry of the nucleus in the SDSS $r$ and $u$ band (see Paper I), the NUV and FUV from \textit{GALEX} and 
\textit{HST} observations, and the unabsorbed power-law X-ray spectrum from \textit{Chandra}. The black dashed line represents a model for a geometrically thin relativistic accretion disc as 
described in the text.}
         \label{fig:multi_SED}
   \end{figure}

\textit{GALEX} observations in the NUV and FUV were taken 2008 October 21 as part of the medium imaging survey when Mrk~1018 was still in the bright phase. Within 100 days it was 
targeted by \textit{Swift} in the X-rays and $U$ band as well as by the Palomar Transient Factory in the $r$ band. The corresponding SED of the nucleus is shown in Fig.~\ref{fig:multi_SED}. The 
SED significantly changes spectral shape in the optical-FUV range when Mrk~1018 was fading, as revealed by the  \textit{HST} observation in combination with quasi-simultaneous $u$ and $r$ band 
photometry (see Paper I).

We modelled the optical/UV radiation as a local black-body radiation from a geometrically thin relativistic accretion disc \citep{Page:1974}. Relativistic effects were included by using the 
\textsc{grtrans}  ray tracing code \citep{Dexter:2016}. For simplicity, we assumed a Schwarzschild BH with fixed $M_\mathrm{BH}=10^8\,M_{\sun}$ (Paper I) and $i=15^\circ$ and fitted for $\dot{M}$ 
at each epoch. The best-fitting models with $\dot{M} \simeq 0.004$ and $0.05\,M_{\sun} \hspace{2pt} \rm yr^{-1}$ are shown in Fig.~\ref{fig:multi_SED}. The simple model provides a satisfactory 
explanation for the change in SED shape, and the inferred factor $\simeq 10$ drop in $\dot{M}$ agrees with the observed decrease in $f_{2-10\,\mathrm{keV}}$.

The good agreement between a static disc model and the spectra with a difference of a factor $\simeq 10$ in luminosity implies that $L \sim T_{\rm eff}^4$ (equivalent to a constant inner radius). 
This relation is frequently seen in BH X-ray binaries \citep[e.g.][]{DavisDB:2006}, but previously has not been found in samples of AGN. 
Their spectra typically peak at $\lambda \simeq 1000$\,\AA\ \citep{Laor:2014}, close to our fit for Mrk~1018 in the bright phase ($\lambda_{\rm peak}\sim1300$\,\AA). The change 
to $\lambda_{\rm peak}\sim$3300\,\AA,  coincident with the decline in luminosity, provides evidence of an optically thick accretion disc.

Cutting the fuel supply of the accretion disc would propagate down to smaller radii on the inflow timescale, $\ga10^2$ years for the optical emitting region in AGN, much longer than in 
Mrk~1018. This timescale is consistent with the disc thermal time, however, and thermal fluctuations can explain the optical/UV variability properties in AGN accretion 
discs \citep{Kelly:2009}. For Mrk~1018 the thermal fluctuation would have to be global, decreasing the temperature across large parts of the disc and not in small patches \citep{Dexter:2011}. 
 
\subsection{What is the nature of the new \Lya\ NAL?}
The most surprising result of our follow-up observations is the appearance of a new associated NAL in \Lya\ that was absent 20 years ago. While it appears natural that this feature is linked to the 
changing-look AGN event of Mrk~1018, it may also be completely unrelated. Below we discuss three potential scenarios for the origin of the NAL, but it may also be something unexpected. 

Given the significant variability of the NAL strength and the radial motion of 700\,\kms\ towards us, the first obvious possibility is that we see an outflow \citep[e.g.][]{Hamann:2013}.  Assuming a 
constant outflow velocity implies a maximum distance of 0.01\,pc, which is rather close to the nucleus. However, the background source may not be 
the accretion disc continuum, but the broad wing in \Lya\ that has larger surface area and might reach up to pc scales as a result of resonant scattering. The NAL could then originate from a 
much faster wind if it is slightly offset from the accretion disc along the line of sight with some inclination. The outflow scenario is attractive because it may be responsible for limiting the gas 
inflow by pushing gas outwards far beyond the outer disc. Disc winds seen in thermal states of BH X-ray binaries \citep[e.g.][]{Ponti:2012} have estimated outflow rates comparable to or larger than 
the inflow rate. The appearing asymmetry towards the blue side of \Ha\ and \Hb\ as reported in Paper I could be a signature of some BLR clouds that are pushed outwards, which still needs to be 
confirmed. 

Another possibility is that the NAL gas is not associated with the accreting SMBH, but orbits a companion SMBH. This could explain the blueshift  with pure gravitational 
motion if the approaching side of clouds around this second SMBH is just moving into our line of sight towards the accretion disc. Since Mrk~1018 is an advanced major merger such a binary SMBH 
scenario is plausible, as speculated in Paper I. The appearing line asymmetry in \Ha\ and \Hb\ might also be signature of the gravitational interactions in a binary SMBH and not only an 
outflow.

Alternatively, the new NAL may be produced by a fast-moving cloud consisting of debris from the major merger. In this scenario the NAL would be disconnected from the accretion rate change. 
However, the speed of the cloud towards our line of sight would require a hyperbolic orbit around the SMBH. 

\section{Conclusions}
Based on follow-up X-ray observations we rule out an obscuring-cloud event as the cause of the change of type again after 30 years, as discovered in McElroy et al. (2016). All observations, in 
particular the optical-UV SED, are consistent with a declining accretion rate of a geometrically thin, optically thick accretion disc. Based on the appearance of a new NAL in \Lya, we speculate 
whether the onset of an outflow or a putative binary SMBH system is driving instabilities in the accretion disc that cause the decline in luminosity. However, the NAL might also be completely 
unrelated to the accretion disc changes. Continuous monitoring from the radio to X-rays is needed to further constrain the nature of the dramatic changes at the heart of the nucleus.

\begin{acknowledgements}
We thank the anonymous referee for very constructive comments and suggestions that significantly helped to improve this letter.
GRT acknowledges support from the NASA through the Einstein Postdoctoral Fellowship Award Number PF-150128, issued by the Chandra X-ray Observatory 
Center, which is operated by the Smithsonian Astrophysical Observatory for and on behalf of NASA under contract NAS8-03060. MK acknowledges support by DFG grant KR 3338/3-1. 
MAPT acknowledges support from the Spanish MINECO through grants AYA2012-38491-C02-02 and AYA2015-63939-C2-1-P. TAD acknowledges support from a Science and Technology Facilities Council Ernest 
Rutherford Fellowship. Parts of this research were conducted by the Australian Research Council Centre of Excellence for All-sky Astrophysics (CAASTRO), through project number CE110001020. 

\end{acknowledgements}


\bibliographystyle{aa}
\bibliography{references}
\end{document}